\documentclass[11pt]{article}
\usepackage[final]{acl}
\usepackage[T1]{fontenc}
\usepackage{times}
\usepackage{latexsym}
\usepackage[utf8]{inputenc}
\usepackage{microtype}
\usepackage{inconsolata}
\usepackage{float}
\usepackage{graphicx} 
\usepackage{booktabs}
\usepackage{amsmath}
\usepackage{eurosym}

\usepackage{xcolor}

\usepackage{listings}
\lstdefinelanguage{json}{
  basicstyle=\ttfamily\small,
  numbers=left,
  numberstyle=\tiny,
  stepnumber=1,
  numbersep=5pt,
  showstringspaces=false,
  breaklines=true,
  frame=single,
  backgroundcolor=\color{lightgray},
  literate=
   *{0}{{{\color{numb}0}}}{1}
    {1}{{{\color{numb}1}}}{1}
    {2}{{{\color{numb}2}}}{1}
    {3}{{{\color{numb}3}}}{1}
    {4}{{{\color{numb}4}}}{1}
    {5}{{{\color{numb}5}}}{1}
    {6}{{{\color{numb}6}}}{1}
    {7}{{{\color{numb}7}}}{1}
    {8}{{{\color{numb}8}}}{1}
    {9}{{{\color{numb}9}}}{1}
    {:}{{{\color{punct}{:}}}}{1}
    {,}{{{\color{punct}{,}}}}{1}
    {"}{{{\color{delim}{"}}}}{1},
}
\definecolor{numb}{rgb}{0.7,0.2,0.2}
\definecolor{punct}{rgb}{0,0,0}
\definecolor{delim}{rgb}{0,0,0.5}
\definecolor{lightgray}{rgb}{0.95,0.95,0.95}

\title{Individual Text Corpora Predict User-Specific Knowledge: Benchmarks of Individualized Knowledge Simulation}

\author{
  Christoph Wigbels$^{1,\dagger}$ \quad
  Ali Abusaleh$^{2,\dagger}$ \quad
  Markus T. Jansen$^{1}$ \AND
  Alexander Mehler$^{2}$ \quad
  Manuel Schaaf$^{2}$ \quad
  Markus J. Hofmann$^{1}$ \\[0.5em]
  $^{\dagger}$ \small{Equal\ contribution} \quad 
  $^{1}$Bergische Universit\"at Wuppertal \quad
  $^{2}$Goethe-Universit\"at Frankfurt \\
  {\small\texttt{\{wigbels, mhofmann, mjansen\}@uni-wuppertal.de}} \\
{\small\texttt{\{a.abusaleh, mehler, manuel.schaaf\}@em.uni-frankfurt.de}}
}

\begin{document}

\maketitle
\begin{abstract}
This study examines whether individual text corpora (ICs) from search histories can be used to simulate individual knowledge. We collected ICs from 316 adults, who answered 36 multiple-choice knowledge items, and compared several large language models (LLMs) on this task, of which only Qwen3-1.7B proved viable. After task-specific fine-tuning via Low-Rank Adaptation (LoRA), Qwen3-1.7B outperformed both participants and
a representative German norm sample on
publicly available items. On non-public
questions, however, the LLM performed worse than our participants, suggesting possible training data contamination for the public questions. When integrating ICs into retrieval-augmented generation to predict individual responses, LLM-participant
Match accuracies significantly exceeded chance, which demonstrates a detectable individual knowledge signal. The probabilities assigned to the participants' answers were, however, low and far below the probability of correct answers, indicating poor calibration toward individual response patterns. Knowledge-gap prediction was suboptimal, though it improved for corpora exceeding five million tokens. We discuss our entropy-based evaluation benchmarks as calibration indices for individualized knowledge simulation.
\end{abstract}

\section{Introduction}

Large language models (LLMs) store substantial amounts of factual knowledge within their parameters. However, this knowledge is static and reflects aggregate patterns in training corpora rather than the knowledge of specific individuals~\cite{petroniLanguageModelsKnowledge2019}. For applications such as adaptive tutoring or personalized diagnostics, it is not sufficient to know what is generally true; instead, a system must model what a particular user knows or does not know to allocate instructional effort effectively~\cite{corbettKnowledgeTracing1995,vanlehnRelativeEffectivenessHuman2011}. The present study addresses this challenge by grounding LLM outputs in individual text corpora (ICs) derived from internet search histories.
Psychologically, individual differences in intellectual abilities are commonly distinguished into fluid intelligence (Gf) and crystallized intelligence (Gc). While Gc is viewed as the ability to use factual and conceptual knowledge (crystallized knowledge; CK), Gf denotes the capacity to solve novel problems independent of Gc~\cite{cattellTheoryFluidCrystallized1963}. Over time, Gf is invested in domain-specific learning, yielding highly individualized knowledge profiles shaped by interests and experience, a component described as the ``dark matter'' of adult intelligence~\cite{ackermanTheoryAdultIntellectual1996, ackermanDomainSpecificKnowledgeDark2000}. Simulating what a particular user knows therefore requires behavioral traces of this individual investment process.
Internet search behavior provides a promising signal: each visited page reflects active engagement with specific content and contributes to a cumulative personal reading history~\cite{marchioniniExploratorySearch2006,dumaisStuffIveSeen2003,teevanInformationReretrieval2007}.
We refer to the factual information encoded in LLM weights as parametric knowledge and use retrieval-augmented generation (RAG) as our computational framework. Standard RAG grounds model outputs in documents retrieved from a shared knowledge base, separating parametric from contextual information~\cite{lewisRetrievalAugmentedGeneration2020,gaoRetrievalAugmentedGeneration2024}.
To evaluate this approach, we collected Google search histories from 316 participants (mean IC size: 3.2~million tokens) and constructed individual corpora via web scraping of visited URLs. Each participant completed 36 multiple-choice knowledge items. To address potential training-data contamination in LLM evaluation~\cite{dengInvestigatingDataContamination2024,zhaoMMLUCFContaminationFree2025}, we distinguished between 12 publicly available BEFKI GC-K items~\cite{schipolowskiBEFKIGCKShort2013}, potentially included in pretraining data, and 24 newly developed items. 
An initial comparison showed that only Qwen3-1.7B reached satisfactory performance; GerPT2 and GPT-2 failed to exceed chance. We therefore evaluate Qwen3-1.7B in three stages: (1) zero-shot parametric knowledge, (2) task-specific fine-tuning via LoRA on 129 training items~\cite{zimnyAntColonyOptimization2024}, and (3) IC-based RAG.
Our evaluation addresses two complementary questions that we keep terminologically distinct throughout the paper. First, we assess factual correctness, that is, whether the selected option matches the keyed correct answer, which we report for the model and for the participants alike. Second, we assess how well the model simulates a specific participant, that is, whether it accurately reproduces the answer this person actually chose, irrespective of whether that answer is correct, and whether it predicts which items this person answered correctly or incorrectly. We reserve the term accuracy for this second, person-oriented perspective and evaluate it using a discrete Match accuracy, a probabilistic log-loss metric, and crystallized-knowledge accuracy (CK-accuracy).

\section{Related Work}

\subsection{Individual Text Corpora and Episodic Retrieval}
While \citet{pennebakerLinguisticStyles1999} provided early correlational
research examining personal diary entries, later work captured
interindividual variability in personality using a common language
model~\cite{eichstaedtClosedOpenVocabulary2021}.
More recent research introduced the IC approach, based on the idea that a text corpus can serve as a sample of a person's experience. A single language model is trained for each person, so that training parallels memory consolidation and inference parallels retrieval~\cite{hofmannIndividualTextCorpora2024}.
\citet{hofmannIndividualCorporaPredict2020} first demonstrated this
approach by constructing ICs from individuals' everyday reading
behavior and training language models directly on these corpora,
showing that such individually trained models predict word fixation
times in an eye-tracking paradigm more accurately than models trained
on standard corpora. \citet{hofmannIndividualTextCorpora2024} then
scaled this approach to web-scale behavioral data, constructing ICs
from Google search histories and showing that IC-derived similarity
measures predict psychological traits such as openness to experience
and intellectual engagement in held-out samples.
Prior studies trained language models directly on ICs, implicitly modeling semantic long-term memory. We instead use RAG to dynamically retrieve relevant IC fragments per query, shifting from consolidated semantic memory to episodic retrieval~\cite{dongTowardsLargeLanguage2025}. This allows us to test whether person-specific corpora support situation-dependent knowledge retrieval.
\subsection{RAG, Knowledge Probing, and Personalized LLMs}
Our work connects to three related research areas in natural language processing (NLP). First, knowledge probing studies have shown that pretrained language models encode substantial factual knowledge that can be accessed through cloze-style queries, while also revealing systematic gaps and inconsistencies in this knowledge~\cite{petroniLanguageModelsKnowledge2019}. 
Second, RAG mitigates these limitations through external retrieval~\cite{lewisRetrievalAugmentedGeneration2020}. Third, recent research on personalized LLMs investigates how user-specific data, such as interaction logs, preferences, or personal documents, can be used to adapt model behavior to individual users~\cite{zhangPersonalizationLargeLanguage2024}. IC-based RAG sits at the intersection of these three lines of work. Like knowledge probing, it is concerned with what a model knows; like RAG, it relies on external retrieval to support generation; and like personalization research, it incorporates user-specific data. Crucially, however, it differs in its evaluation target: instead of optimizing for general factual correctness or user preference, we evaluate whether the model can reproduce individual knowledge. This shifts the focus from generic knowledge augmentation to the simulation of person-specific cognitive states.

\section{Methods}
\subsection{Participants and Knowledge Assessment}
The final sample comprised 316 adults (71.4\% female; 65.6\% aged 18--25, 27.3\% aged 26--35) with diverse educational backgrounds (63.0\% Abitur, 16.9\% university degree). Participants were recruited via university flyers, the SONA system, the PsyWeb panel\footnote{\url{https://psyweb.uni-muenster.de}}, and personal recruitment. Eligibility required fluency in German, active use of a Google account for at least one year, and participation in a 60--90-minute online survey covering knowledge, personality, interests, fluid intelligence, and demographic information. Participants received either course credits or 18\,\euro{}. Participants were excluded if they did not provide a valid search history file, if their IC contained fewer than 2{,}500 stemmed word types, or if they failed all control questions.
Crystallized knowledge was assessed using 36 multiple-choice items. These included the 12 items from the BEFKI GC-K short scale~\cite{schipolowskiBEFKIGCKShort2013} and 24 new items. In addition, 129 items \citep{zimnyAntColonyOptimization2024}\footnote{\url{https://osf.io/u68nk/files}} were used for task-specific fine-tuning and are distinct from the 36 evaluation items. Each item consisted of a question stem and four answer options, with exactly one correct answer; an example item is shown in Appendix~\ref{app:osq_item}.
  
\subsection{Individual Text Corpora and Web Scraping}
ICs were constructed from participants' search histories. Participants exported their data via Google Takeout; extracted URLs were then scraped using a two-pass pipeline (HTTP extraction followed by JavaScript rendering) with rotating proxies and PDF support. A cleaning filter removed boilerplate content and non-German text (sentences with fewer than three German stop words were excluded).

We next evaluated how pretrained language models perform on the task before adapting them to the specific question format and integrating participant-specific corpora via RAG.

\subsection{Task-Specific Fine-Tuning of Pretrained Language Models}

We investigated the capability of LLMs to process and answer multiple-choice questions within a structured evaluation framework. 
Our methodology comprised three primary phases: (1) establishing baseline performance through zero-shot evaluation,
(2) adapting the models to the specific question format using Low-Rank Adaptation (LoRA)~\cite{hu:et:al:2021:lora} rather than full-parameter fine-tuning, and (3) simulating participant-specific knowledge by integrating personal corpora via RAG during the question-answering process.

\subsection{Baseline and Task-Specific Fine-Tuning}

\paragraph{Baseline Evaluation}  
We first evaluated several language models on the multiple-choice task in a zero-shot setting to establish a baseline. The evaluated models included GPT-2~\cite{radford:et:al:2019language}, GerPT2~\cite{Minixhofer:GerPT2:2020}, and Qwen3-1.7B~\cite{qwen3:2025}. Performance correctness was calculated based on an exact match between the model's predicted option\footnote{Given the small size of the test dataset, manual post-processing and extraction of the answers were sufficient for evaluation.} and the ground-truth answer.  

\paragraph{Task-Specific Fine-Tuning} 

To adapt the model to the structured multiple-choice format while keeping computational costs low, we applied parameter-efficient fine-tuning using LoRA. Rather than updating all model parameters, LoRA introduces trainable low-rank adapter matrices into selected layers while keeping the original weights frozen.

The fine-tuning dataset comprised 129 multiple-choice items formatted identically to the evaluation items. Each prompt included a question stem and four answer options, with the correct answer encoded as a letter (A--D). The prompt format enforced a standardized output structure compatible with downstream evaluation (see Listing~\ref{lst:promptFormat}).

\begin{lstlisting}[float, language=python, caption={Prompt for QA Fine-tuning}, label={lst:promptFormat}, breaklines=true]
Please answer with one of the options in the bracket. 
Write the answer
in between <answer></answer>.

### Question:
{question text with options A-D}

### Response:
<answer>
{correct letter}. {correct text}
</answer>
\end{lstlisting}

Fine-tuning was performed with a LoRA rank of $r=8$, a scaling factor $\alpha=16$, and a dropout rate of $0.05$ for regularization. Adapters were injected into multiple projection layers (\texttt{q\_proj}, \texttt{k\_proj}, \texttt{v\_proj}, \texttt{o\_proj}, \texttt{gate\_proj}, \texttt{up\_proj}, and \texttt{down\_proj}) to maximize adaptation capacity.
Training used the 32-bit Paged AdamW optimizer (\texttt{paged\_adamw\_32bit}) with a learning rate of $2 \times 10^{-4}$. Models were trained for 10 epochs with a per-device batch size of 1 and 2 gradient accumulation steps, along with 10 warmup steps to stabilize training.

After adapting the model to the task format, we incorporated participant-specific information via RAG.

\subsection{RAG Pipeline}

Each corpus was segmented into chunks of 400 tokens with an overlap of 50 tokens to preserve contextual continuity. These chunks were embedded using the \texttt{nomic-embed-text} model~\cite{Nussbaum:et:al:2025:nomic}, producing 768-dimensional vector representations, and stored in a Qdrant vector database~\cite{qdrantQdrantVector}.
At inference time, the system retrieves the top $k = 5$ most semantically relevant chunks for a given question. These retrieved text fragments are then incorporated into the model prompt, grounding the generated response in the participant's personal corpus.

\begin{figure}[!t]
  \centering
  \includegraphics[width=0.9\columnwidth]{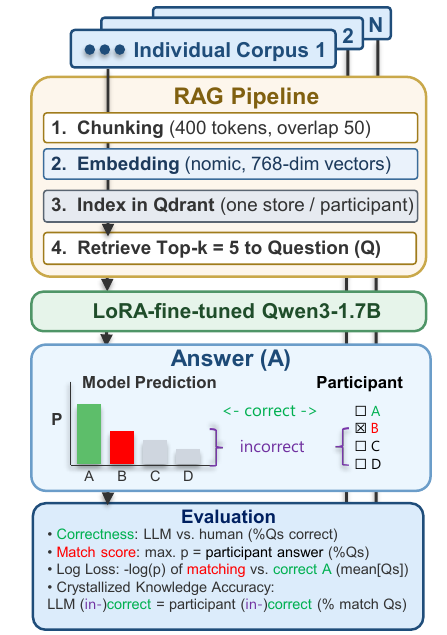}
  \caption{Study Overview: Language Modeling Pipeline and Evaluation}
  \label{fig:overview}
\end{figure}

These modeling steps yield both discrete predictions and full probability distributions over answer options. We evaluate model performance using complementary metrics that capture both correctness and probabilistic calibration (see Figure~\ref{fig:overview}).

\paragraph{Inference Prompt Design} 

While fine-tuning, the model was conditioned on the multiple-choice format (Listing~\ref{lst:promptFormat}). At evaluation time, the prompt is augmented with retrieved participant context and an explicit instruction to rely on it (Listing~\ref{lst:evalPrompt}).
The instruction directs the model to answer \emph{based only} on the provided search-history context, rather than on its parametric knowledge, with a fallback clause that permits reliance on the participant's demonstrated interests when the answer is not directly retrievable.

\begin{lstlisting}[float, language=python, caption={Inference prompt combining Context and Question}, label={lst:evalPrompt}, breaklines=true]
You are a specialized assistant. 
You MUST answer the question 
based ONLY on the provided search
history context. If the answer is 
not in the context, use your best 
judgment based on the participant
interests shown in the history.

### SEARCH HISTORY CONTEXT:
{retrieved participant search history}
## QUESTION:
{question text}
A. {option A}
B. {option B}
C. {option C}
D. {option D}
### Response:
<answer>
\end{lstlisting}

\subsection{Evaluation}

We evaluated model performance along four complementary dimensions.

\paragraph{Answer correctness.}
We first measured standard multiple-choice factual knowledge correctness by comparing the model's predicted option with the keyed correct answer.

\paragraph{Match accuracy.}
To assess how well the RAG model reproduces individual responses, we computed the Match accuracy, defined as the proportion of items for which the model's most likely option matches the participant's choice.

\paragraph{Log loss.}
To evaluate the full predicted response distribution, we computed multiclass log loss. For each item $i$, the model produces a probability vector $\mathbf{p}_i = (p_{i,1}, \dots, p_{i,4})$, and the participant's response is represented as a one-hot vector $\mathbf{y}_i$. This allows us to interpret the model's uncertainty in information-theoretic terms: log loss measures how much probability mass the model assigns to the observed response and is equivalent to the cross-entropy 
\begin{align}
H(\mathbf{y}_i, \mathbf{p}_i) = - \sum_{j=1}^{4} y_{i,j} \log p_{i,j},
\end{align}
which, for one-hot targets, is equivalent to the Kullback--Leibler divergence
\begin{align}
D_{\mathrm{KL}}(\mathbf{y}_i \,\|\, \mathbf{p}_i)
= \sum_{j=1}^{4} y_{i,j} \log \frac{y_{i,j}}{p_{i,j}}.
\end{align}
Averaging across $N$ items yields the log-loss score,
\begin{align}
\mathrm{LogLoss}
= -\frac{1}{N} \sum_{i=1}^{N} \log p_{i,j_i^{*}},
\end{align}
where $j_i^{*}$ denotes the option chosen by the participant. Log loss is measured in natural units of information (nats). Lower values indicate better alignment between predicted probabilities and observed responses. For instance, the chance-level baseline corresponds to a uniform distribution ($-\log 0.25 = 1.39$ nats).

\paragraph{CK-accuracy.}
While the Match accuracy asks whether the maximum probability answer of the LLM is the participant's selected answer, we were also interested in a measure that asks whether the LLM can accurately predict whether a participant answered an item correctly or incorrectly. Therefore, we evaluated knowledge-gap prediction using crystallized-knowledge accuracy (CK-accuracy), defined as the proportion of items for which the model accurately predicts whether a participant answered correctly or incorrectly.

\begin{table}[!ht]
\centering
\begin{tabular}{lcc}
\toprule
\textbf{Model} & \textbf{Zero-shot} & \textbf{Fine-Tuned} \\
\midrule
GerPT2 & 0\% & 4\% \\
GPT-2 & 0\% & 22.7\% \\
Qwen3-1.7B & 81.1\% & 100\% \\
\bottomrule
\end{tabular}
\caption{LLM Selection Based on Correctness}
\label{tab:selection}
\end{table}

All metrics were averaged across participants and compared to their respective baselines using one-sample \emph{t}-tests.

\section{Results}

\begin{table*}[!t]
  \centering
  \small
    \begin{tabular}{llcccccc}
      \toprule
      Metric & Item set & $M$ ($SD$) & Baseline & \emph{t} & $p$ & $d$ & 95\% CI \\
      \midrule
      Participant correctness
        & All (36)      & .64 (.14) & .67\textsuperscript{a} & $-3.79$  & $< .001$ & $-0.21$ & $[-.044, -.014]$ \\
        & Core (12)     & .63 (.16) & .81\textsuperscript{a} & $-20.12$ & $< .001$ & $-1.13$ & $[-.204, -.168]$ \\
        & Extended (24) & .65 (.14) & .60\textsuperscript{a} & $6.10$   & $< .001$ & $0.34$  & $[.034, .066]$   \\
      \addlinespace
      Norm correctness\textsuperscript{e}
        & Core (12)     & .59 (.22) & .81\textsuperscript{a} & $-33.67$ & $< .001$ & $-1.00$ & $[-.233, -.207]$ \\
      \addlinespace
      Match Accuracy
        & All (36)      & .31 (.06) & .25\textsuperscript{b} & $18.84$  & $< .001$ & $1.06$  & $[.057, .070]$   \\
        & Core (12)     & .36 (.10) & .25\textsuperscript{b} & $19.50$  & $< .001$ & $1.10$  & $[.102, .125]$   \\
        & Extended (24) & .29 (.07) & .25\textsuperscript{b} & $9.52$   & $< .001$ & $0.54$  & $[.031, .047]$   \\
      \addlinespace
      Log loss: match (nats)
        & All (36)      & 6.29 (0.72) & 1.39\textsuperscript{c} & $121.06$ & $< .001$ & $6.81$ & $[4.820, 4.978]$ \\
        & Core (12)     & 6.26 (1.19) & 1.39\textsuperscript{c} & $72.69$  & $< .001$ & $4.09$ & $[4.742, 5.005]$ \\
        & Extended (24) & 6.30 (0.82) & 1.39\textsuperscript{c} & $105.98$ & $< .001$ & $5.96$ & $[4.821, 5.003]$ \\
      \addlinespace
      Log loss: correct (nats)
        & All (36)      & 1.56 (0.26) & 1.39\textsuperscript{c} & $12.22$  & $< .001$ & $0.69$  & $[.148, .205]$   \\
        & Core (12)     & 0.99 (0.29) & 1.39\textsuperscript{c} & $-23.64$ & $< .001$ & $-1.33$ & $[-.425, -.359]$ \\
        & Extended (24) & 1.86 (0.35) & 1.39\textsuperscript{c} & $23.96$  & $< .001$ & $1.35$  & $[.435, .512]$   \\
      \addlinespace
      CK-accuracy
        & All (36)      & .54 (.08) & .64\textsuperscript{d} & $-22.67$ & $< .001$ & $-1.28$ & $[-.107, -.090]$ \\
        & Core (12)     & .54 (.14) & .63\textsuperscript{d} & $-11.41$ & $< .001$ & $-0.64$ & $[-.103, -.073]$ \\
        & Extended (24) & .55 (.09) & .65\textsuperscript{d} & $-21.48$ & $< .001$ & $-1.21$ & $[-.112, -.094]$ \\
      \bottomrule
    \end{tabular}
    \caption{One-Sample \emph{t}-tests Comparing Evaluation Metrics Against Their Respective Baselines}
    \label{tab:results}
    \small\emph{Note.} $N = 316$ for all metrics except norm correctness. Core = BEFKI GC-K short form \citep{schipolowskiBEFKIGCKShort2013}; Extended = newly developed items not publicly available. Values in parentheses are standard deviations. 95\% CIs refer to the mean difference $M - \text{Baseline}$. All \emph{t}-tests use $df = 315$ except norm correctness, which uses $df = 1{,}133$. Log loss: correct = log loss computed against the keyed correct answer rather than the participant's chosen option.
    \textsuperscript{a}LLM correctness (item-level proportion correct).
    \textsuperscript{b}Four-option chance level.
    \textsuperscript{c}Uniform distribution: $-\ln(0.25) = 1.39$~nats.
    \textsuperscript{d}Majority-class baseline (overall participant correctness per item set).
    \textsuperscript{e}Norm correctness = correctness for a nationally representative German norm sample ($N = 1{,}134$; \citealp{schipolowskiBEFKIGCKShort2013}).
\end{table*}

\subsection{LLM Selection}
Models were evaluated on a held-out test set of 22 multiple-choice items \citep{zimnyAntColonyOptimization2024} that were not included in the fine-tuning corpus and are distinct from the 36 evaluation items used in the main analysis (Table~\ref{tab:selection}).
The results indicate that, while smaller models (GPT-2, GerPT2) showed modest improvements~\cite{Minixhofer:GerPT2:2020,radford:et:al:2019language}, the Qwen3-1.7B model reached 100\% correctness on the held-out set after fine-tuning~\cite{qwen3:2025}.
Correctness before fine-tuning ranged from 0\% to 81.1\%; the gains after fine-tuning demonstrate the impact of task-specific adaptation.
However, in this initial selection set, the correct option was always A, so a model could in principle have reached a perfect score simply by always producing the letter A, reflecting label bias rather than genuine knowledge. We therefore (i) treated this comparison only as a preliminary viability check for model choice and (ii) retrained Qwen3-1.7B for the final evaluation on items with randomized correct-answer positions, so that no constant-response strategy could inflate its scores.
\subsection{LLM Versus Human Answer \texorpdfstring{Correctness}{Correctness}}

\begin{figure}[t]
  \centering
  \includegraphics[width=\columnwidth]{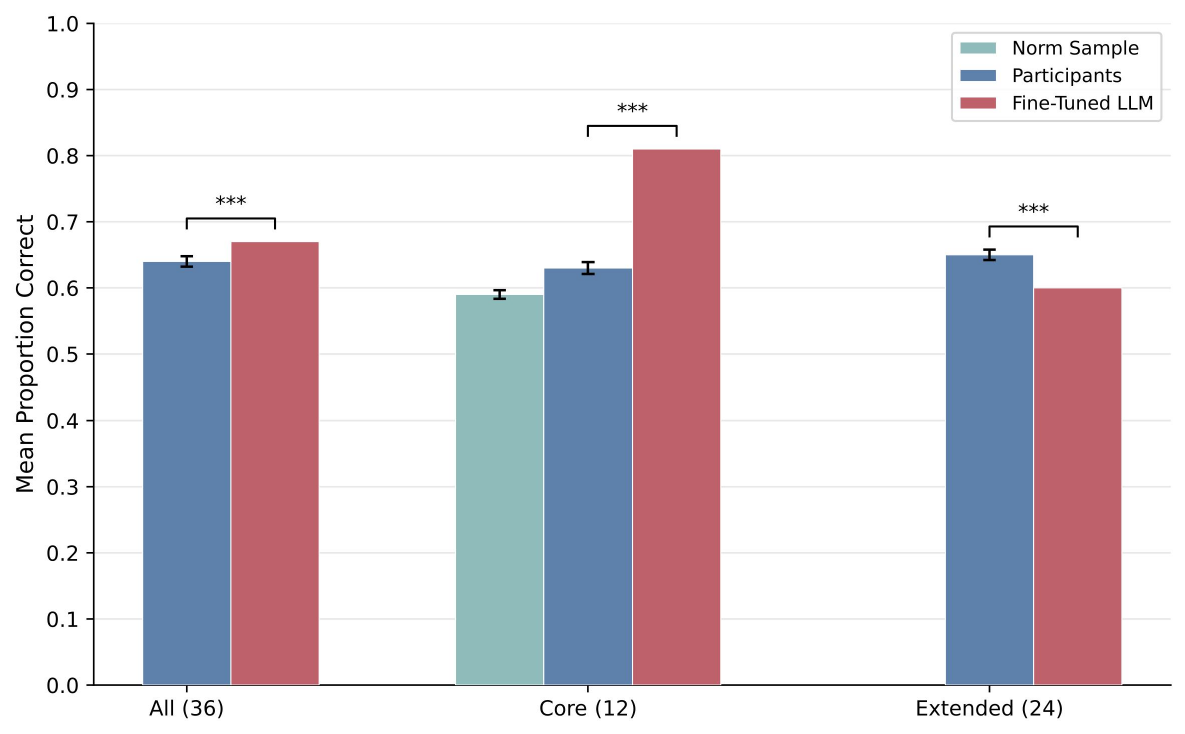}
  \caption{Mean Proportion Correct for the Representative 
    Norm Sample, Study Participants, and the Fine-Tuned 
    LLM by Item Set}
  \label{fig:accuracy}
  \begin{minipage}{\columnwidth}
    \small
    \emph{Note.} Error bars represent $\pm 1$~\emph{SE}. 
    $N = 316$; norm data from a nationally representative 
    German sample ($N = 1{,}134$; 
    \citealp{schipolowskiBEFKIGCKShort2013}), available 
    for core items only. LLM correctness is a fixed item-level 
    proportion (no error bar). $^{***}p < .001$.
  \end{minipage}
\end{figure}

Participant correctness (mean proportion correct per participant) was compared to the fixed item-level correctness of the fine-tuned Qwen3-1.7B model (Figure~\ref{fig:accuracy}). One-sample \emph{t}-tests were used to test whether the participant mean differed from the constant LLM correctness (Table~\ref{tab:results}).
Across all 36 items, the LLM achieved higher correctness than participants. However, this effect was confined to the 12 BEFKI GC-K core items, where the model substantially outperformed the participant sample ($81\%$ vs.\ $63\%$). On the 24 extended items, participants outperformed the LLM ($65\%$ vs.\ $60\%$).
To contextualize these results, we compared both groups to a nationally representative German norm sample ($N = 1{,}134$; \citealp{schipolowskiBEFKIGCKShort2013}). On the core items, the norm sample scored $.59$ ($SD = .22$), which is below our participant sample ($.63$) and the LLM ($.81$).

\subsection{Individualized Knowledge Simulation}
IC-based RAG Match accuracies and log loss were tested against chance ($0.25$ for Match accuracy; $1.39$~nats for log loss) using one-sample \emph{t}-tests (see Table~\ref{tab:results}).
Match accuracies significantly exceeded chance across all item sets, reaching an overall mean of $.31$ against the $.25$ chance level ($\Delta = +.06$, $d = 1.06$), with the strongest effect for core items ($.36$) and a moderate effect for extended items ($.29$).
However, this improvement in discrete matching did not translate into well-calibrated probabilistic predictions. Log loss was consistently higher than chance across all item sets, indicating that the model assigned low average probabilities to the responses actually chosen by participants. This discrepancy reflects a systematic pattern: while the model often selected the same answer as the participant, it did so with highly concentrated probability distributions.
The low mean normalized entropy ($.10$) further indicates that the model failed to represent uncertainty across alternative options.

A complementary analysis computing log loss against the keyed correct answer revealed a dissociation: for core BEFKI items, log loss fell below baseline, whereas for extended items it was clearly above baseline. 

Finally, when evaluating whether the model can predict which items a participant answered correctly or incorrectly, CK-accuracy fell below the majority-class baseline across all item sets (see Table~\ref{tab:results}), indicating that the model did not reliably distinguish between items that participants knew and those they did not.

\subsection{Corpus Size as a Moderator}

\begin{figure}[htbp]
  \centering
  \includegraphics[width=\columnwidth]{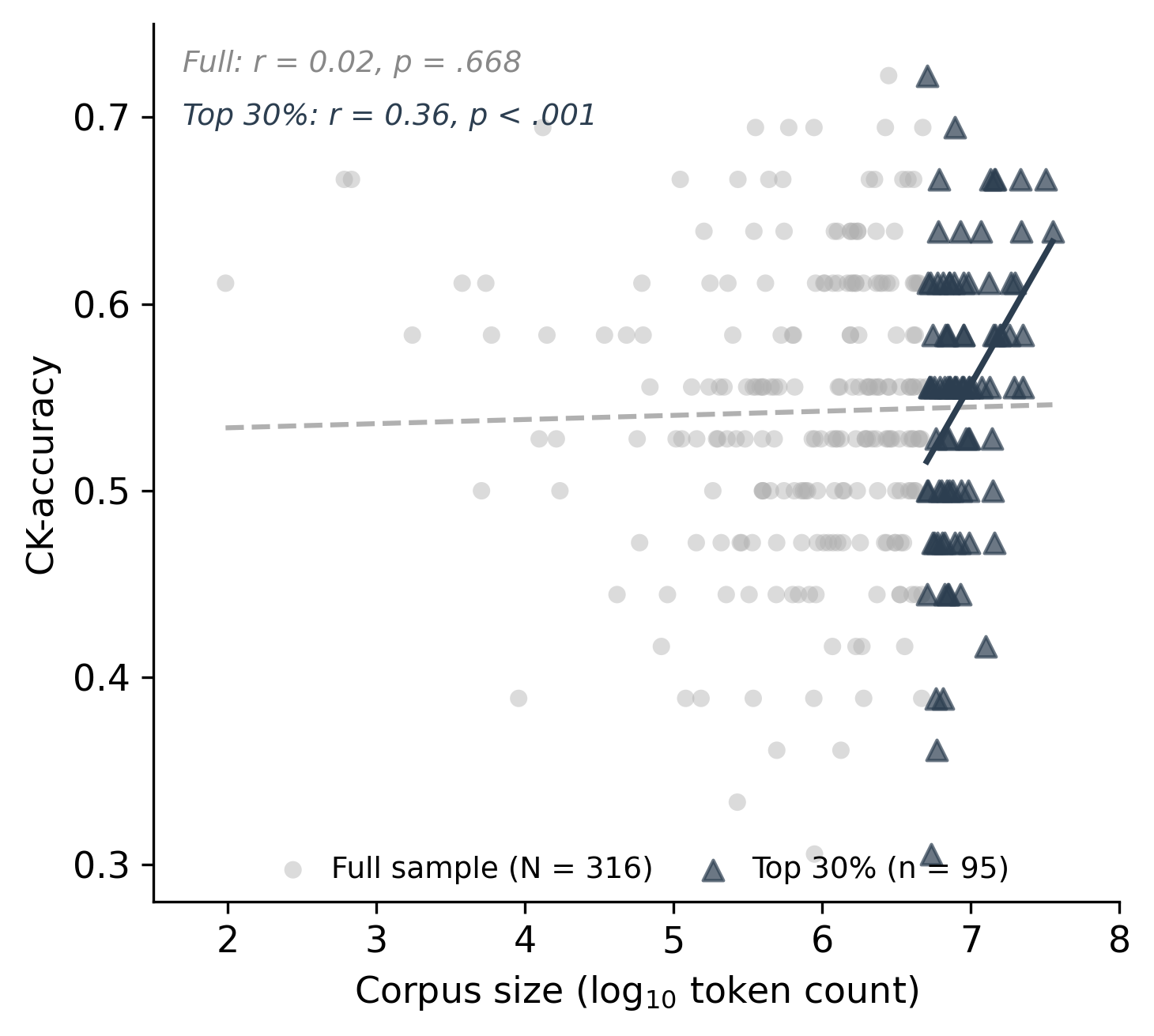}
  \caption{Corpus Size and CK-accuracy for the Full Sample and the Top-30\% Subsample}
  \label{fig:corpus}
  \begin{minipage}{\columnwidth}
    \small
    \emph{Note.} Pearson correlation between corpus size ($\log_{10}$ token count) and CK-accuracy. Gray circles (dashed regression line) = full sample ($N = 316$); dark triangles (solid regression line) = top-30\% subsample ($n = 95$; $\geq 5.02$ million tokens). $^{*}p < .05$. $^{***}p < .001$.
  \end{minipage}
\end{figure}

We further explored the association between corpus size ($\log_{10}$ token count) and evaluation metrics with partial correlations, thus controlling for participant correctness. Corpus size was not significantly associated with any evaluation metric (all $p$s $> .13$). However, a different pattern emerged among participants with larger corpora. In the top-30\% subsample ($n = 95$; $\geq 5.02$~million tokens), corpus size was positively associated with CK-accuracy, $r(93) = .36$, $p < .001$ (see Figure~\ref{fig:corpus})
and with participant correctness, $r(93) = .25$, $p = .013$. The association between corpus size and CK-accuracy remained significant after controlling for participant correctness, partial $r(92) = .28$, $p = .006$. These findings indicate that prediction quality improves once a sufficient corpus size threshold is reached.

\section{Discussion}

This study examined whether ICs derived from Google search histories can serve as a proxy for person-specific crystallized knowledge. As reported above, only Qwen3-1.7B proved viable after task-specific fine-tuning; all subsequent analyses relied exclusively on this model.

A notable dissociation emerged when comparing model performance across different item types. On the publicly available BEFKI core items, the LLM outperformed human participants ($81\%$ vs.\ $63\%$ correctness), whereas on the newly developed extended items, it slightly underperformed ($60\%$ vs.\ $65\%$ correctness). This pattern is consistent with the problem of training-data contamination in standardized assessments, and it simultaneously demonstrates that human crystallized intelligence still surpasses parametric world knowledge when confronted with genuinely novel problems. In this regard, the results underscore the urgent need for contamination-controlled benchmarks in LLM evaluation, as has been argued in recent work \citep[e.g.,][]{jacoviBenchmarkContamination2023}. Performance on public items likely reflects parametric retrieval rather than genuine inference, which suggests that standard benchmark scores may substantially overestimate a model's actual reasoning capacity. The 18-percentage-point advantage on the public BEFKI core items, compared with the model's underperformance on the contamination-controlled extended items, illustrates this concern concretely and is consistent with the view that novel, previously unseen evaluation sets are necessary when claims about model reasoning are to be assessed.

Regarding the central question of individualized knowledge simulation, IC-based RAG yielded Match accuracies, that is, agreement with the answers the participants actually chose, consistently above the chance level of $.25$ (overall $.31$, $\Delta = +.06$, $d = 1.06$). To our knowledge, this result constitutes the first empirical demonstration that person-specific text corpora derived from search histories carry a detectable signal of individual knowledge. 
The large effect size primarily reflects low between-person variance rather than a substantial practical gain. Moreover, the log-loss analysis reveals a fundamental limitation that correctness metrics alone would obscure: the observed log-loss values indicate that the model fails to reflect the distribution of incorrect options a participant might actually consider, though it predicts the correct option with extreme confidence.

This dissociation points to a critical optimization gap: future individualized models must shift from answer-oriented to person-oriented training, for instance by incorporating participants' actual response distributions during fine-tuning. 
Furthermore, CK-accuracy fell below the majority-class baseline ($.54$ vs.\ $.64$), indicating that the model does not yet capture individual variations in knowledge gaps. This asymmetry, whereby the model performs above chance for what participants know but fails for what they do not know, offers at least two interpretations. First, the structure of the data itself plays a role: search histories record moments of active information seeking rather than failed recall, such that gaps manifest as silence within the IC. Second, a theoretical explanation rooted in PPIK (Process, Personality, Interests, and Knowledge) theory \citep{ackermanTheoryAdultIntellectual1996,ackermanDomainSpecificKnowledgeDark2000} suggests that digital behavior primarily reflects areas of sustained intellectual engagement; the absence of search activity is therefore only a weak indicator of missing knowledge, as a person may possess knowledge acquired through non-digital channels. 
Crucially, the predictive signal scaled with data volume: only in the top-30\% subsample ($\geq 5.02$~million tokens) did corpus size significantly predict CK-accuracy, even after controlling for participant knowledge. This suggests that the approach is currently data-limited rather than fundamentally flawed, and the threshold at approximately five million tokens establishes a concrete design parameter for future personalized RAG systems.

\section{Conclusions and Outlook}
This study demonstrates that ICs from search histories carry a measurable signal of person-specific crystallized knowledge when combined with RAG. Match accuracies establish a proof of concept, but log-loss analysis identifies calibration as the primary bottleneck. The corpus-size effect locates the current bottleneck in data volume rather than in the approach itself.
Together, our benchmarks allow for a clear developmental trajectory: shifting from answer-oriented to person-oriented optimization, enriching ICs with behavioral metadata such as dwell time and re-visitation frequency, and ensuring sufficient corpus size above the approximately five-million-token threshold identified here.
In particular, minimizing log loss against participants' actual choices could directly incentivize calibrated probability estimates.
From a cognitive perspective, we interpret RAG-based retrieval as an episodic memory retrieval process \citep{dongTowardsLargeLanguage2025}, though individuals likely also differ in semantic long-term memory. Our benchmarks may also help investigate the inability of LLMs to forget pretraining knowledge \citep{jangKnowledgeUnlearning2023, yaoMachineUnlearningPretrained2024, blancoJusticiaDigitalForgetting2025}. Given the corpus-size effect, future work should include English-language websites.

A first and immediate extension addresses the scope of the present evaluation, which rested on a single model. Model size is only one of several levers for individualizing language models \citep{zhangPersonalizationLargeLanguage2024}. Because the poor calibration we observed coincided with the strong parametric factual knowledge of Qwen3-1.7B, smaller models such as Qwen3-0.6B, with less memorized world knowledge to fall back on, are a natural next comparison, consistent with evidence that smaller language models can align more closely with human knowledge distributions than
larger ones \citep{heyueya2024psychometric}. Such small models may leave more room for the individual signal to shape the answer distribution, which would speak to whether the calibration gap is tied to model scale. In parallel, the person-oriented objective sketched above can be realized through per-participant fine-tuning that writes each individual corpus directly into the model weights with a weight-decomposed low-rank adapter \citep{liu2024dora}, optimizing the
model to reproduce a specific person rather than the keyed correct answer. We are currently pursuing both directions, and report an exploratory comparison of further models under the identical pipeline in Appendix~\ref{app:model_comparison}.
A further direction concerns the integration of interests as a complementary individualizing signal. Because knowledge gaps remain largely invisible to IC-based retrieval, as reflected in the below-baseline CK-accuracy, the corpus alone says little about what a person does not know. Interests, which the present survey already assessed with the FIFI-K, a short inventory of leisure interests \citep{nikstatFIFIK2018}, offer a principled if broad proxy: within the PPIK framework, sustained interest drives the acquisition of domain knowledge, so that low interest in a domain is itself informative about likely gaps. As leisure interests capture only one facet of the interests that drive knowledge acquisition, a domain-specific or epistemic interest measure would sharpen this signal further. Combining the retrieved corpus content with an explicit interest profile could thus improve knowledge-gap prediction precisely where the corpus fails to provide a signal. Ultimately, successful individualized simulation could enable tutoring agents to select instructional texts that minimize the log-loss gap between model predictions and correct answers.

\section*{Limitations}

Notwithstanding the previously mentioned findings, several limitations must be critically noted. First, the sample was predominantly young, female, and highly educated (65.6\% aged 18--25; 71.4\% female; 63.0\% Abitur), which may limit the generalizability of the results to demographic groups whose search behavior and knowledge profiles differ systematically. Second, the study is language-specific at this stage: all knowledge items were administered in German, all ICs were filtered to retain German-language content, and the stopword-based filter may have excluded relevant multilingual material. Also, it remains unclear whether the approach generalizes to other languages or to multilingual users whose search histories span multiple languages.
Third, the construction of the ICs relied exclusively on Google search histories. Although this source provides a scalable form of data donation, it captures only one outlet of intellectual engagement and excludes other potentially informative digital traces, such as e-book annotations or YouTube watch history. The latter is also accessible via Google Takeout and could be incorporated in future work through auto-generated captions.
Fourth, the mapping from the IC to what a person actually knows could be indirect. A visited page may index interest and information seeking rather than secured knowledge, so that the presence of a topic in the corpus does not imply that the person mastered it, also reading a text does not guarantee that its content was understood or retained. Beyond this, web search itself can function as a form of external, or transactive, memory: as people come to expect continuous online access to information, they tend to remember where to find it rather than the content itself \citep{sparrowGoogleEffectsMemory2011}. A search may therefore not reflect internalized knowledge at all, and it can instead represent precisely the content a person has offloaded to the web, which further weakens the corpus-knowledge link. The IC is therefore a proxy for exposure and engagement rather than for knowledge itself, complementing the reverse point discussed above. In line with this, the overlap between the corpora and the tested items is uneven and follows the searchability of a topic rather than its curricular importance: everyday knowledge that people routinely look up, for instance the Zugspitze, the legal form GmbH, the founding of the German Reich, or Nietzsche, tends to be well represented, whereas canonical school knowledge such as mitosis or the mitochondrion appears far less often. Notably, coverage was not higher for the publicly available BEFKI core items than for the non-public extended items, so that coverage does not simply track an item's difficulty or public availability.
Fifth, only a single model architecture (Qwen3-1.7B) and one retrieval configuration were evaluated in the main RAG setting. The model's high answer correctness was accompanied by high overconfidence, which likely contributed to the poor log-loss performance. This highlights a central limitation of the present approach: strong item-level correctness does not necessarily imply well-calibrated simulation of individual response distributions. Because no alternative model sizes, calibration procedures, or retrieval strategies were compared, it remains unclear whether the observed trade-off is specific to the present implementation or reflects a more general limitation of IC-based answer simulation.

\section*{Ethical Considerations and Data Availability}
The data collection for the overarching research project was approved by the ethics committee of Bergische Universität Wuppertal. All participants provided informed consent, including explicit agreement to share their Google search history data for scientific purposes. We invested considerable effort into anonymizing the individual text corpora. The ICs do not contain person-identifying information, raw URLs, or precise timestamps; only coarse timestamps relative to assessment time are retained. Because ICs represent a subset of the information collected during commercial web tracking, participant identifiability should be lower than with standard tracking data \citep{deusserBrowsingUnicityLimits2020}.
As web search data are already used extensively for commercial purposes, we consider it an ethical necessity to lead an open scientific discussion about the possibilities and limitations of such data for psychological research. In contrast to commercial objectives, our work aims to improve individualized psychodiagnostics and adaptive tutoring.
Nevertheless, the risk of re-identification from large personal text corpora cannot be fully excluded. For this reason, the ICs are not publicly released at this time. Instead, we plan to provide them through the GESIS data archive under controlled access, so that secondary analyses remain restricted to scientific purposes covered by the participants' informed consent. Because the residual risk of re-identification should be measured rather than merely assumed, we further plan a shared task on person identifiability. This shared task rests on the same safeguards as the present study, that is, explicit consent to data donation, the anonymization described above, and controlled access for registered teams, so that the diagnostic benefit of quantifying identifiability outweighs the additional risk. Please contact us if you are interested in participating.

\section*{Acknowledgments}
Funded by grants from the Deutsche Forschungsgemeinschaft (DFG, German Research
 Foundation) 
-- project numbers \href{https://gepris.dfg.de/project/539645652}{539645652} and \href{https://gepris.dfg.de/project/539634240}{539634240} -- within the priority programme
``New Data Spaces for the Social Sciences'' (SPP 2431).

\bibliography{references}

\appendix

\section{\texorpdfstring{Example Items}{Example Items}}
\label{app:osq_item}

The 129 fine-tuning items and the 22 held-out selection items were drawn from the openly available general-knowledge item pool of \citet{zimnyAntColonyOptimization2024}.\footnote{\url{https://osf.io/u68nk/}} The following item (\texttt{mus073}, solution rate $.49$ in the pool) illustrates the format. All items were administered in German; the English translation in brackets is taken from the pool documentation, and the correct option is printed in bold.

\begin{samepage}
\begin{quote}
Welches Instrument gehört nicht zu der Gruppe der Holzblasinstrumente? \\
{[}Which instrument does not belong to the group of woodwind instruments?{]} \\[0.5em]
$\Box$~\textbf{Waldhorn} {[}\textbf{french horn}{]} \\
$\Box$~Querflöte {[}transverse flute{]} \\
$\Box$~Oboe {[}oboe{]} \\
$\Box$~Fagott {[}bassoon{]}
\end{quote}
\end{samepage}

The 12 BEFKI GC-K core items follow the same four-option format and are documented in \citet{schipolowskiBEFKIGCKShort2013}, whereas the 24 extended items are deliberately not published, to keep them out of public text collections as the contamination control, in line with standard practice for psychological test material. The following core item, reproduced from the public BEFKI GC-K documentation in the GESIS instrument archive ZIS\footnote{\url{https://doi.org/10.6102/zis220}}, illustrates the format; as no official English translation exists, the bracketed translation is our own.

\begin{samepage}
\begin{quote}
Wozu dient die Mitose? \\
{[}What is the function of mitosis?{]} \\[0.5em]
$\Box$~Stoffwechselregulation {[}metabolic regulation{]} \\
$\Box$~Fortpflanzung {[}reproduction{]} \\
$\Box$~Bildung von Keimzellen {[}formation of germ cells{]} \\
$\Box$~\textbf{Zellvermehrung bei Wachstumsvorgängen} {[}\textbf{cell proliferation during growth}{]}
\end{quote}
\end{samepage}

\section{\texorpdfstring{Comparison of Further Models}{Comparison of Further Models}}

\label{app:model_comparison}
To probe whether the calibration gap identified in the main analysis is tied to model scale, we evaluated a set of additional models under the identical IC-based RAG pipeline, retrieval configuration, and probability extraction (Table~\ref{tab:model_comparison}). While larger models tend to show more correct answers, the fine-grained log-loss metric indicates that smaller models should be more suitable for individualized knowledge simulation.

\begin{table}[!h]  
  \centering
  \footnotesize
  \setlength{\tabcolsep}{3pt}
  \begin{tabular}{lcccccc}
    \toprule
    \textbf{Model} & \textbf{Params} & \textbf{Corr.} & \textbf{MA} & \textbf{LL:M} & \textbf{LL:C} & \textbf{CK} \\
    \midrule
    Qwen3-0.6B\textsuperscript{a} & 0.6B & .46 & .33 & 2.24 & 1.82 & .43 \\ 
    LFM2-700M & 0.7B & .65 & .52 & 2.22 & 1.10 & .62 \\
    Qwen3-1.7B\textsuperscript{a} & 1.7B & .67 & .31 & 6.29 & 1.56 & .54 \\
    Qwen3-4B-Instruct-2507 & 4B & .83 & .57 & 5.47 & 1.07 & .60 \\
    Mistral-7B-v0.3 & 7B & .88 & .60 & 3.07 & 0.74 & .63 \\
    \bottomrule
  \end{tabular}
   \caption{Model Comparison Under the Identical IC-Based RAG Pipeline}
  \label{tab:model_comparison}
  
  \begin{minipage}{\columnwidth}
    \footnotesize\emph{Note.} Corr. = Correctness; MA = Match accuracy; LL:M = Log loss: match; LL:C = Log loss: correct; CK = CK-accuracy. All rows except the Qwen models are unadapted base checkpoints evaluated under the identical retrieval configuration and probability extraction, without task-specific fine-tuning; $N = 484$--$498$; values are means over participant$\times$item pairs in the with-context condition. Chance level: $.25$ (MA), $1.39$~nats (LL); majority-class baseline for CK: $.66$. \textsuperscript{a}Fine-tuned on the QA task ($N = 316$).
  \end{minipage}
\end{table}

Across the tested models the individual context barely shifts answer behavior: for Mistral-7B, correctness ($.875$ vs.\ $.861$) and match accuracy ($.596$ vs.\ $.609$) are near-identical with and without retrieval, and the shift is smallest for the model with the strongest parametric knowledge. This is expected if the answer distribution is dominated by a sharp parametric prior that the retrieved context cannot outweigh. Our target, however, is not correctness but accuracy in the sense of reproducing a participant's answer behavior, and a strong prior maximises the former while leaving little room for the latter. The bottleneck is therefore not model size but how the individual signal enters the model: retrieval adds it as inference-time evidence that competes with the prior and loses. A consequent next step, sketched in the Conclusions, is to write each participant's corpus into the model weights, where the prior itself resides, with per-participant adaptation of smaller models as a secondary lever for a less dominant prior.

\end{document}